# *Scalable surface area characterisation by electrokinetic analysis of complex anion adsorption*

Dorian A.H. Hanaor [(1)], Maliheh Ghadiri [(2)], Wojciech Chrzanowski [(2)], Yixiang Gan [(1)]

*(1) School of Civil Engineering, University of Sydney, NSW 2006, Australia*
*(2) Faculty of Pharmacy, University of Sydney, NSW 2006, Australia*

**Abstract**:

By means of in-situ electrokinetic assessment of aqueous particles in conjunction with the addition of anionic adsorbates, we develop and examine a new approach for the scalable characterisation of specific accessible surface area of particles in water. For alumina powders of differing morphology in mildly acidic aqueous suspensions, effective surface charge was modified by carboxylate anion adsorption through the incremental addition of oxalic and citric acids. The observed zeta potential variation as a function of proportional reagent additive was found to exhibit the inverse hyperbolic sine type behaviour predicted to arise from monolayer adsorption following the Grahame-Langmuir model. Through parameter optimisation by reverse problem solving, the zeta potential shift with relative adsorbate addition revealed a near-linear correlation of a defined surface-area-dependent parameter with the conventionally measured surface area values of the powders, demonstrating that the proposed analytical framework is applicable for the in-situ surface area characterisation of aqueous particulate matter. The investigated methods have advantages over some conventional surface analysis techniques owing to their direct applicability in aqueous environments at ambient temperatures and the ability to modify analysis scales by variation of adsorption cross-section.

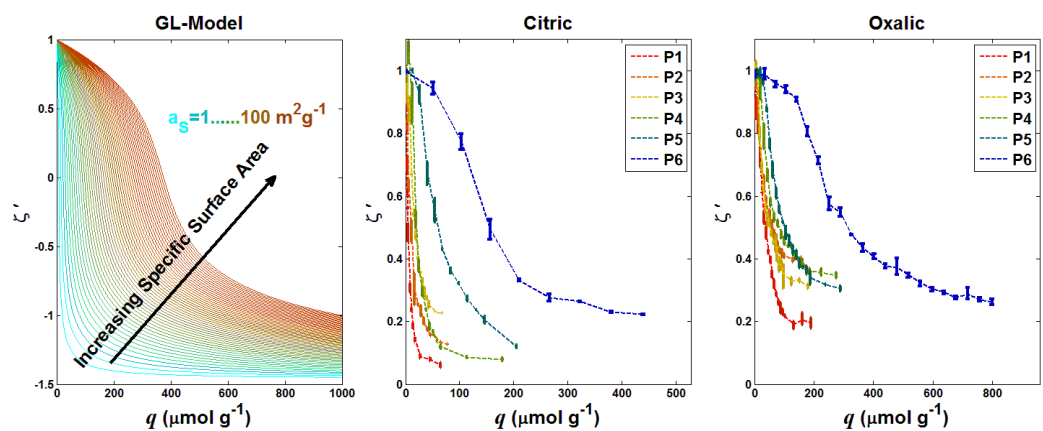

## 1. Introduction:

The ability to meaningfully characterise particle interface structures in aqueous media is of importance in a range of high value industrial processes and applications including catalysis[1], pharmaceutics[2], water treatment[3] and across the broader field of chemical engineering.

In particulate material, as for natural surfaces in general, a unique value for specific surface area, in terms of area per unit mass, cannot be categorically defined. In similarity to the well-known coastline paradox, this results from the scale variance of surface structures [4-5]. Thus over the past decades for the assessment of surface driven material functionality we commonly speak of the gas-accessible surface area, most frequently measured by $N_2$ adsorption in conjunction with BET or Langmuir type isotherm interpretation for multi-layer or mono-layer gas adsorption [6-9]. While improved sensitivity and accuracy is achievable through the adsorption of heavier gases (Krypton or Argon), such methods nonetheless suffer from known limitations with respect to measurement scale and conditions [10-11].

Conventional methods, such as BET, have the clear advantage of facilitating standardised comparative analyses using purpose-built commercially available analytical apparatus. However, there are known drawbacks to the use of such tools. Specifically gas adsorption methods are limited with respect to: (i) measurement scale – $N_2/O_2/Kr/Ar$ exhibit molecular adsorption cross-sections in the range 0.14-0.23 nm$^2$ [12] (ii) measurement temperature – BET analysis is most commonly undertaken at cryogenic temperatures (e.g. 77K for $N_2$) (iii) measurement environment – such methods are generally applied to dry powder. The aforementioned scale variance of surface structures means that the assessment of surface area at a constant measurement resolution is problematic. Furthermore, in many applications including catalysis, environmental remediation and in chemical engineering in general, particulate materials are applied in an aqueous environment at ambient or high temperatures, thus motivating the adoption of surface area characterisation tools that can be applied in analogous conditions, with the aim of conducting target application relevant interface characterisation.

Among alternative adsorption based surface area analysis methods put forward over recent decades aqueous and organic suspension based methods feature prominently. Typically the analysis of specific surface area by adsorption in liquid media involves the selection of a material- and application-appropriate adsorbate compound, or 'molecular probe', and the intermittent analysis of an indicative parameter to characterize the presence of residual free adsorbate in inter-particle fluid[13]. This is typically achieved *ex-situ* using calorimetric, spectroscopic, titration-based or visual inspection of inter particle fluids [14-17]. This type of surface characterisation is encumbered by the need for parallelised analyses and the limitation to systems involving complete or near complete adsorption.

The importance of the hierarchical or fractal nature of particle interfaces with surrounding media has resulted in an increasingly wide range of adsorption-based studies addressing the description and measurement of surface area scale variance in particulate materials. Such research efforts were pioneered by studies by Avnir et. al. in a series of publications in the 1980s and 90s [5, 18-22]. Conventional nitrogen adsorption isotherms can interpreted to yield information regarding surface fractality using the Frenkel-Halsey-Hill Theory [20, 23]. This method has limitations and its application to systems of unknown surface area is problematic. Further methods to probe scale variance of small aqueous particles, exhibiting high refractive indices, include laser light scattering interpreted using Rayleigh-Gans-Debye theory [23-24]. More recently, electrochemical approaches to characterising roughness and fractal surface structures in electrodes have been reported using cyclic voltammetry, double layer capacitance analysis and diffusion limited current measurement [25-27]. While not utilised to gauge accessible surface area, these studies highlight the applicability of using multi-ionic interactions for scalable surface analyses.

Although the formation of multilayers of polyelectrolytes has been reported [28], the adsorption of complex ions, i.e. molecular ionic species, at aqueous particle surfaces is best described by Langmuir isotherms (Type-I), appropriate due to the electro-sterically limited quasi-monolayer type adsorption exhibited [29-30]. Saturation is approached with increasing adsorbate surface density as the result of electrostatic repulsion of charged species limiting further surface ligation.

The electrokinetic behaviour exhibited by suspended inorganic particles is known to vary with the adsorption of surfactant molecules to particle surfaces. Recent studies reported the variation of zeta-potential (the electric potential at the shear surface between a particle and its suspending media) with adsorption of carboxylate anions to surfaces of $TiO_2$ and $ZrO_2$ [31-32]. These studies found that the double layer behaviour observed in suspensions was governed by parameters of adsorbate size and particle surface area.

In the present work we investigate the merit of electrokinetic analyses for the direct assessment of adsorbate-accessible surface area of aqueous granular materials in a recirculating suspension. By introducing a new methodology to interpret adsorption isotherms through indicative zeta-potential variation we gauge the appropriateness of electrokinetic analysis in the analysis of surface structure and particle-reagent interactions.

## 2. Methodology

Solute ionic adsorbates ligating to particle surfaces in suspension form a quasi-monolayer, the density of which exhibits an electro-steric limit towards steady state conditions, governed by the size and charge of adsorbates. Consequently this process can be described by a Langmuir type adsorption isotherm relating fractional surface coverage to adsorbate concentration [30, 33]. Fractional surface coverage $\theta_f \in [0,1]$, is defined as the ratio of the areal density of surface adsorbed molecules $N_S$ to the total number of effective surface sites per unit area $N_{tot}$. For adsorption from solution to particles with a given total surface area this is expressed following the Langmuir form as:

$$\theta_f = \frac{N_S}{N_{tot}} = \frac{\kappa C}{1 + \kappa C} \qquad (1a)$$

Here $C$ corresponds to the volumetric concentration of adsorbate in the system, and the coefficient $\kappa$ corresponds to the ratio of adsorption/desorption for a given adsorbate/adsorbent pair such that:

$$\kappa = \frac{[AS]}{[A][S]} \qquad (1b)$$

Where $[A]$ is the concentration of adsorbate in solution and $[AS]$ and $[S]$ represent the surface densities of occupied and unoccupied sites on the adsorbent particles. $[A]$ is inversely proportional to the system volume while $[S]$ is proportional to the total number of effective surface sites and thus in monolayer chemisorption for a constant adsorbent mass ($m_a$), the coefficient $\kappa$ is proportional to the system volume $V$ and inversely proportional to the available adsorbent surface area $A_s$, which in turn is the product of specific surface area ($a_s$) and $m_a$. An expression to account for specific surface area scaling can be written as

$$\kappa = \kappa' V A_s^{-1} \ , \quad \theta_f = \frac{\kappa' C V a_s^{-1} m_a^{-1}}{1 + \kappa' C V a_s^{-1} m_a^{-1}} = \left( \frac{\kappa'^{-1} a_s}{q} + 1 \right)^{-1}$$

(2)

where $q=CV/m_a$ corresponds to the quantity of adsorbate per gram of adsorbent (in mol/gram).

We can define a surface area dependant adsorption coefficient $K$ [mol/g]:

$$K \propto a_s \ \text{ such that } \ K = \kappa'^{-1} a_s \quad \text{(3a)}$$

Thus the expression given in Eq. 2. simplifies to the expression

$$\theta_f = \frac{1}{Kq^{-1} + 1} \qquad \text{(3b)}$$

The adsorption of complex ions is associated with an increase in surface charge density $\sigma$ with relation to the charge density of particle surfaces in the absence of adsorbate, $\sigma_0$. The density of surface charge in turn governs electrokinetic parameters [34].

On the basis of the Gouy Chapman theory, following the Grahame equation form, the Stern potential $\psi_\delta$ (at the plane of adsorbed species) and zeta potential $\zeta$ are generally related to the apparent surface charge, $\sigma$, of a suspended particle through an inverse hyperbolic sine relationship of the type given in Eq. 4.

$$\zeta = M_1 \text{arcsinh}(M_2 \sigma) \qquad \text{(4)}$$

The parameter $\sigma$ may refer to true surface charge of the solid phase or of observed Stern-layer charge and accordingly $M_1$ is a system constant related to variables of temperature, permittivity and counter-ion concentration and speciation, while $M_2$ is a constant dependant on temperature and shear plane separation [35-37].

It has been found that the adsorption of complex ions to particle surfaces in aqueous suspension manifests in a shift of zeta potential with relative adsorbate concentration following a sigmoidal form with electrokinetic parameters tending towards a quasi-steady state as adsorption capacity is approached [31]. For adsorption of anionic species to suspended particles exhibiting initially positive zeta potential values, the shift in zeta potential is expected to exhibit the relationship shown in Eq. 5 as a function of fractional surface coverage $\theta$. The parameter $\sigma'_m$ corresponds to the net shift in observed double-layer charge (within the sphere defined by the slipping plane) brought about by monolayer adsorption (occupation of all effective sites) at a surface exhibiting an initial charge of $\sigma_0$ in the absence of adsorption.

$$\zeta = M_1 \text{arcsinh}\left[ M2\left(\sigma_0 - \theta\sigma'_m\right)\right] \qquad \text{(5)}$$

As the adsorption of ionic species in aqueous solution manifests in the formation of a charged quasi-monolayer, the generalised relationship given by the Grahame-Langmuir model describing the dependency of potential to adsorbate concentration can be written as:

$$\zeta = M_1 \text{arcsinh}\left[ M_2\left(\sigma_0 - \frac{\sigma'_m}{\left(1 + Kq^{-1}\right)}\right)\right] \quad \text{(6)}$$

For the anionic adsorption to particles exhibiting an initially positive zeta potential, as studied in the present work, the proportional zeta potential ($\zeta'$, taken relative to the initial value of $\zeta_{(\theta=0)}$ ) exhibits the trend shown in Figure 1. as a function of surface coverage $\theta$. Here plots are shown for increasing (negative) adsorbate charge.

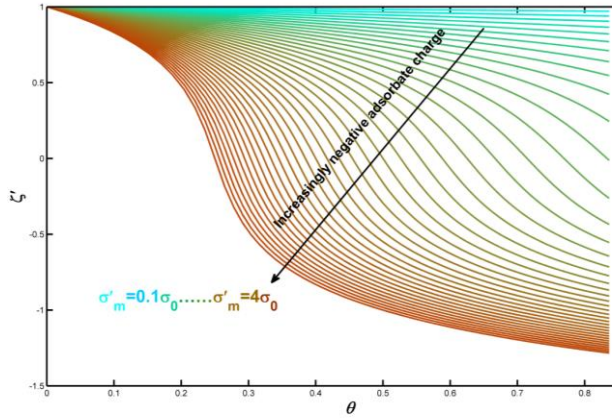

**Figure 1. Relative zeta potential as a function of fractional surface coverage for anionic adsorption for a range of different adsorbate anion charge values following Eq. 5.**

In similarity to fractional surface coverage, we can define a readily measureable quantity of fractional zeta potential shift $\beta \in [0,1]$ such that:

$$\beta = \phi \left| \frac{\zeta - \zeta_0}{\zeta_s - \zeta_0} \right| \qquad (7)$$

Under given conditions $\beta$ represents the double layer modification at conditions of zeta potential $\zeta$ relative to the surface-saturated state. Here the value of $\zeta_0$ corresponds to the zeta potential of suspended particles at the given pH level in the absence of adsorbate and $\zeta_s$ represents the zeta potential exhibited by suspended particles under conditions equivalent to monolayer coverage ($\zeta_s = \zeta_{(\theta=1)}$). $\phi$ is a correction factor included to account for the shift of solution parameters, namely pH and ionic strength, of interparticle fluid (outside the Stern layer) with adsorbate addition. For a system of low solids loading where the ionic strength and pH are assumed to remain sufficiently stable so as not to impart significant $\zeta$ manipulation, we accept a value of $\phi = 1$ (and hence $\beta_{(\theta=1)} = 1$).

$$\beta = \left| \frac{\operatorname{arcsinh}\left[ M_2 \left( \sigma_0 - \frac{\sigma'_m}{(1 + Kq^{-1})} \right) \right] - \operatorname{arcsinh}[M_2 \sigma_0]}{\operatorname{arcsinh}[M_2(\sigma_0 - \sigma'_m)] - \operatorname{arcsinh}[M_2 \sigma_0]} \right|$$

(8)

Consequently $\beta$ is expected to vary following the plot shown in Figure 2. It can be seen that increasing specific surface area, linearly correlated to $K$, manifests in a shift of the sigmoidal relationship.

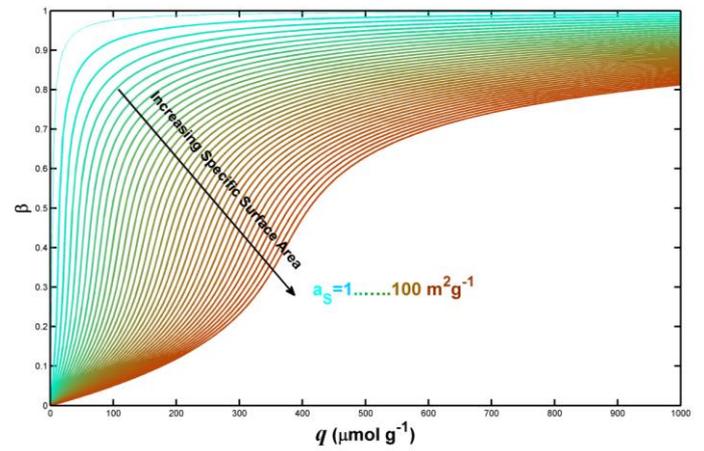

**Figure 2. Fractional zeta potential shift $\beta \in [0,1]$, as s function of additive concentration for a given system volume.**

By zeta potential analysis in conjunction with adsorbate addition, the parameter $\beta$ and its variation can be measured experimentally *in situ* in suspensions or slurries containing particles of unknown surface structure to gauge accessible surface area per unit mass at tuneable scales.

## 3. Experimental procedures:

In order to evaluate the electrokinetic based surface characterisation of particles in aqueous suspension, calcined and ground high purity alumina powders ($Al_2O_3$, Baikowski, >99.9% purity) of varying particle size and specific surface area were chosen as characteristic adsorbent materials. Powder characteristics are summarised in Table 1. Morphology of powders was assessed by Scanning Electron Microscopy analysis at 5 kV acceleration by means of a Zeiss-Ultra SEM. Zeta potential measurements were achieved using a Malvern Nano ZS analyser with an automated peristaltic additive dispensing system with suspension recirculation. This apparatus utilises Phase Analysis Light Scattering (PALS) to assess the electrophoretic mobility of particles and thus facilitate the measurement of zeta potential in-situ (in the recirculating suspension). Samples were suspended in deionized water to give 10 ml suspensions with solids loadings of 0.1 wt%. In order to impart positive initial zeta potential values and adequate deflocculation, suspensions were adjusted to pH= 4 with the dropwise addition of diluted HCl. Citric and Oxalic acids (99%, Univar) were used as anionic adsorbates, incrementally added in the form of dilute aqueous solutions to stirring alumina suspensions by means of automated dispensing. For each additive increment, three EK measurements were taken with one minute intervals separating the measurements.

**Table 1. Characteristics of alumina powders used in the present work**

| Sample Designation | Supplier designation | Milling method | $Al_2O_3$ Phases | BET surface area ($m^2g^{-1}$) | Agglomerate size (nm) |
|---|---|---|---|---|---|
| P1 | CR1 | Jet milled | α | 3 | 1100 |
| P2 | CR6 | Jet milled | α | 6 | 600 |
| P3 | SMA6 | Ball milled | α | 7 | 300 |
| P4 | CR15 | Jet milled | 90% α  10%γ | 15 | 400 |
| P5 | CR30F | Jet milled | 80% α  20%γ | 26 | 400 |
| P6 | CR125 | Jet milled | γ | 105 | 300 |

## 4. Results and Discussion

SEM micrographs of the $Al_2O_3$ powders used are shown in Figure 3. A typical hierarchical microstructure is observed with hard agglomerates consisting of finer primary particles, with the size of primary particles and agglomerate fractality resulting in an increasing specific surface area from P1 to P6.

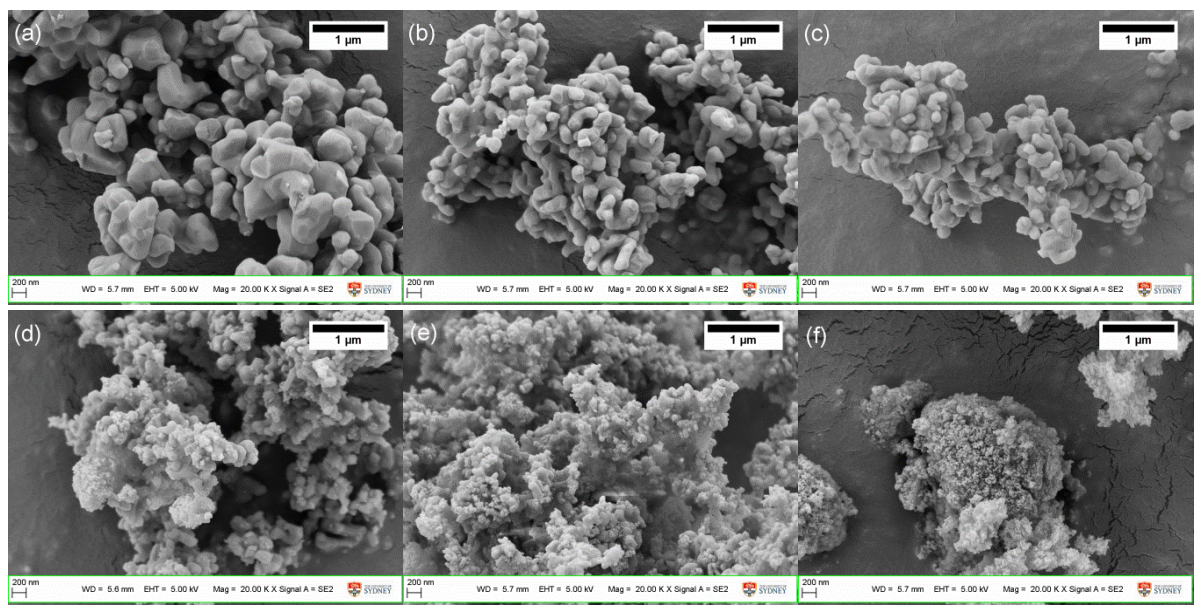

**Figure 3. SEM micrographs of $Al_2O_3$ powders used (a)-(f) corresponding to P1-P6 in Table 1.**

In similarity to previous observations from $ZrO_2$ and $TiO_2$ suspensions [31-32], the addition of dilute citric and oxalic acids to pH=4 alumina suspensions brings about a measurable shift of the electrokinetic properties of suspended particles through the surface adsorption of carboxylate anions. This is shown in Figure 4, where zeta potential is plotted against the relative adsorbate addition (in proportion to the mass of particles). Here dashed lines show mean values while vertical bars show the data spread. Zeta potential was found to vary from an initial value of ~45-60 mV in the HCl adjusted pH = 4 suspension to a final value in the range 0-10 mV subsequent to significant carboxylate addition. From repeated measurement at various concentrations it was established that for the studied systems, the observable change in surface charge resulting from carboxylate adsorption reaches equilibrium in less than two minutes. For this reason, measurements involved an equilibration time under stirring before repeated zeta-potential analyses were carried out.

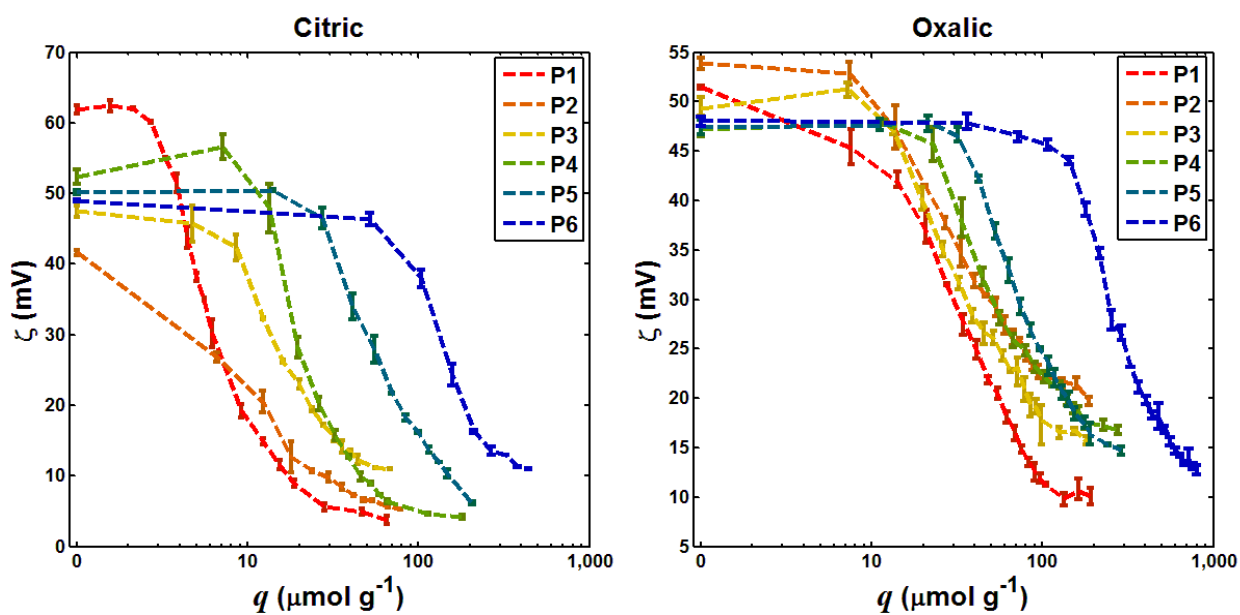

**Figure 4. Variation of Zeta potential with additive ratio for powders P1-P6 with (a) citric acid and (b) oxalic acid adsorbates. Vertical bars show data range at each measurement point.**

It can be seen that curves resulting from the use of oxalic acid are shifted to higher relative concentration values (with respect to solids mass) in comparison with the results from citric acid addition owing to the smaller adsorption cross section of oxalate anions relative to citrate. Although being substrate and speciation dependant, these cross sections on oxide materials are reported varyingly in the region of ~0.6 and ~1.4 $nm^2$ for oxalate and citrate anions respectively [38-41]. Similarly, with increasing specific surface area (P1...P6) changes in electrokinetic behaviour are observed at higher additive levels. Figure 5 shows the measured relative shift in zeta potential ($\zeta' = \zeta/\zeta_0$) in comparison to that predicted from the Grahame-Langmuir relationship from Eq. 6.

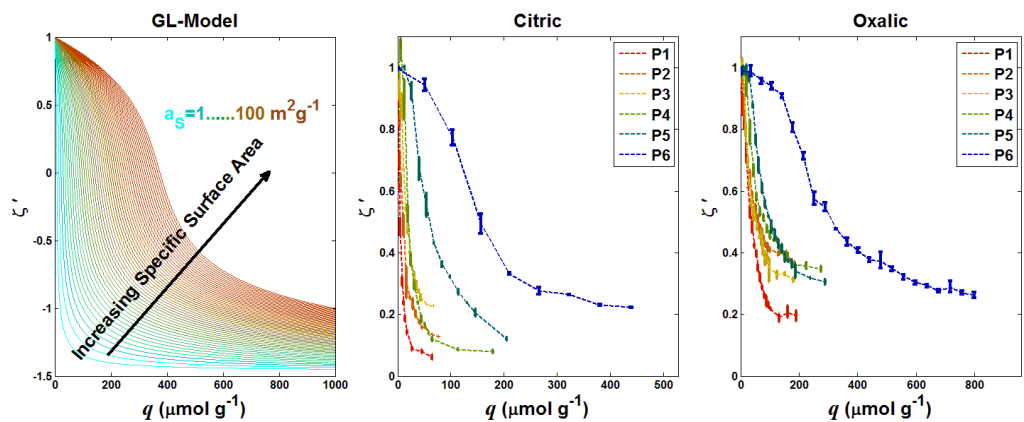

**Figure 5. Relative zeta potential shifts resulting from oxalate and citrate adsorption compared to the Grahame Langmuir model.**

In addition to facilitating initial deflocculation, the pre-acidification of suspensions ensured that the observed electrokinetic behaviour was mediated by reagent interface interactions rather than by changes to suspension pH or ionic strength. This is further illustrated in Figure 6, showing values for $\zeta$, pH and conductivity during the addition of carboxylic reagents to representative suspensions of P4. The slight decrease in pH would typically be expected to bring about an increase in $\zeta$ values rather than a decrease, while the moderate increase in suspension conductivity would too not be expected to significantly vary the electrokinetic behaviour of particles. An initial increase in pH is found in similar systems as the result of the displacement of surface hydroxyls with carboxylate adsorption[31].

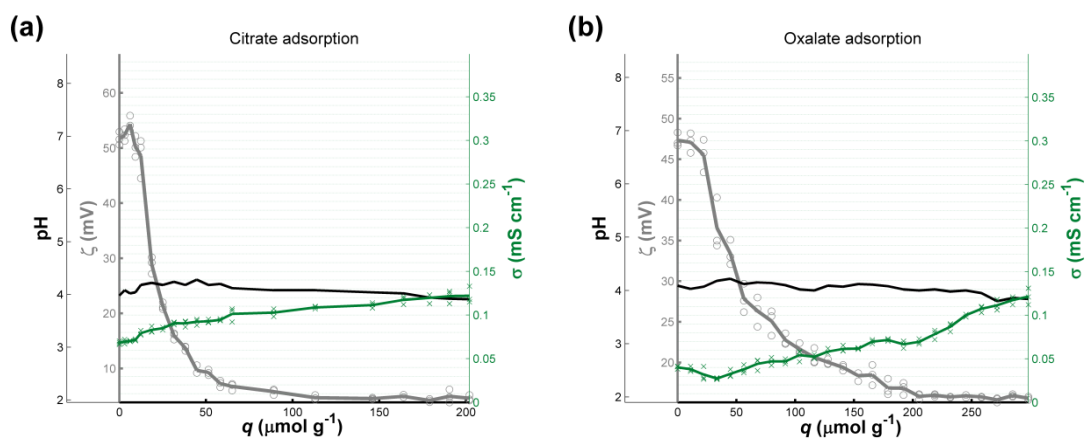

**Figure 6. Variation of $\zeta$, pH and conductivity in P4 suspensions with (a) citrate and (b) oxalate adsorption.**

The measureable differences in the indicative electrokinetic behaviour seen between substrate powders of different surface area and adsorbates of different size, as shown here, demonstrate the applicability of electrokinetic probing for in-situ surface area assessments of aqueous particles. Towards this end, parameter fitting by reverse problem solving is carried out in order to quantitatively evaluate this relationship. While BET measurements are generally limited to the gas accessible specific surface area relative to $N_2$ at 77 K (although other gases are utilisable), the present approach gives an additional degree of freedom enabling scale specific surface characterisation at room temperature in aqueous media by using complex ionic adsorbates of varied effective size.

**Identification of model parameters:**

Averaged data for the relative shift in zeta-potential was interpreted using a parallelised least squares type fitting process to determine optimised values for the model parameters in Eq. 8. Simultaneous multiple parameter optimisations for all particle types were carried out separately for results from both oxalic and citric additives using an unbounded Levenberg Marquardt algorithm. Tolerance was set as $1 \times 10^{-14}$ and convergence was achieved after ~300 iterations using centred finite differences for curves for both oxalic and citric adsorbates. In this manner an optimised $K$ value for each paring of adsorbent/adsorbate was determined, while optimal values for $M_2\sigma_0$ and $M_2\sigma_m'$ were evaluated for each of the two adsorbate reagents optimised to yield the best combined fitting across all 6 curves in each dataset. The results of data fitting for each of the six powders are shown in Figure 7.

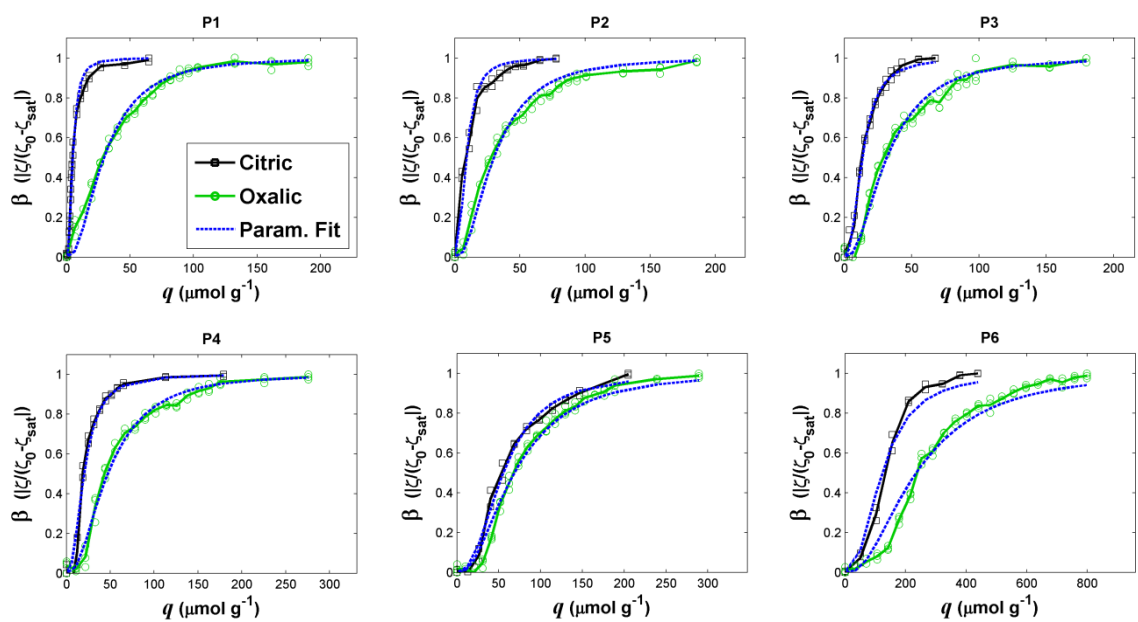

**Figure 7. Parameter optimisation for fractional $\zeta$ shift as a function of proportional additive ratio.**

The close agreement of the fitted parameters and the experimental data suggest that under the conditions employed, adsorption behaviour followed approximately the monolayer form of the Grahame-Langmuir model. Alternative adsorption models, as may be appropriate for protein/polymer multilayer adsorption or other non-type-I forms [42-44], can facilely be incorporated into the present approach through suitable substitution or adaptation of the Langmuir isotherm form with a more appropriate model for the evaluation of the $\theta/C$ relationship in the numerical framework used for parameter fitting.

From the data fitting carried out it was determined that for the adsorption of citrate anions the ratio of net negative change in apparent Stern plane charge relative to initial conditions, $\dfrac{\sigma_m'}{\sigma_0}$, was 2.04, while for oxalate adsorption this value was optimised at 1.9. These values are indicative of charge density formed by monolayer adsorption of these carboxylate anions and are dependant on parameters of ion speciation, adsorption /desorption rates and maximum surface coverage density.

For a given system the optimised value for the parameter $K$ ($K_{fit}$) is expected to exhibit linear dependence on scale-specific effective surface area, following Eq. 3. While no true single value of specific surface area can be defined, examining the validity of this model requires comparison with a standardised surface area metrology method. Therefore, to further assess the applicability of the current methods towards surface area analysis we examine the correlation between K values and the conventional N₂ adsorption isotherm determined surface area. As shown in Figure 8, this comparison of $K_{fit}$ values against BET derived surface area for the 6 powder types shows a trend supporting the linear correlation of fitted K values with specific surface area. Deviations from linear behaviour are expected to result from a combination of (a) fundamental discrepancies between room temperature anionic adsorption in aqueous media and gas adsorption at low temperatures and (b) experimental uncertainties in establishing and fitting the curves of electrokinetic variation. It is likely that the first parameter is of greater relevance for powders of higher surface area while the latter issue plays an important role for substrate materials of lower specific surface area, which exhibit greater sensitivity in their EK behaviour.

$K_{fit}$ values are found to be approximately 1.6-2.0 times higher in results from incremental oxalic acid addition relative those obtained using citric additive, this is consistent with the larger adsorption cross section of the citrate anion as reported for various oxide substrates. It should be noted that as with gas adsorption methods[12], the precise adsorptive cross section in terms of nm² per molecule is contingent also on the substrate material and consequent density of exchangeable surface groups. For this reason, in contrast to comparative analysis, meaningful quantitative analyses require a system specific calibration to establish a reference point for a known surface area under given conditions. Furthermore, the monolayer density is affected by the adsorbate speciation, which in turn is influenced by pH, meaning significant pH fluctuations would be detrimental to the accuracy of the analysis. For the conditions utilised here (pH=4) oxalate is expected to exhibit a speciation of 57% AH⁻ and 43% A²⁻ , while citrate is expected to speciate following 76% AH₂⁻ 13%AH²⁻ and 11% AH₃ (where A represents the fully deprotonated molecule)[31].

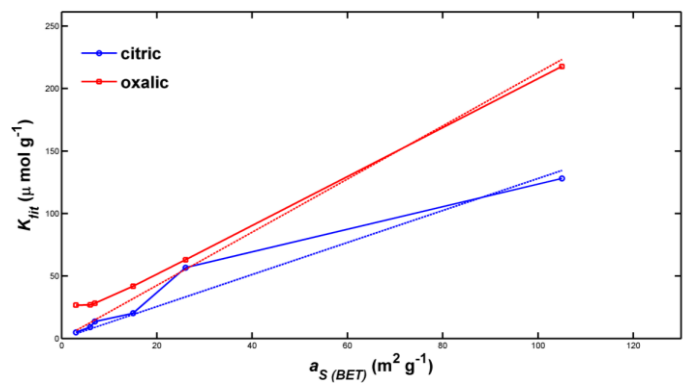

**Figure 8. Optimised values for surface-area normalised adsorption coefficient *K* plotted against BET measured surface area values.**

The approximate linearity of the $K_{fit}$ /$a_{S(BET)}$ relationship is indicative of the applicability of electrokinetic analysis for the quantitative or comparative evaluation of surface area. Correlation coefficients of $R^2_{citric}$=0.946 and $R^2_{oxalic}$=0.965 were obtained for this relationship fitted to pass through the origin $K_{fit}$ $(a_S=0)=0$. Owing to the fundamental discrepancies between BET methods and the adsorption methods explored here, it may be appropriate to additionally examine the correlation of results with surface area obtained by ex-situ analysis of adsorption (e.g. spectroscopic analysis of supernatant fluids) or by microanalytical methods (AFM, SEM or TEM).

The slope of $K$/$a_s$ ($\kappa'^{-1}$) in units of μmol m⁻² is proportional to the density of monolayer adsorption. For the fitted data, The slopes of 2.13 μmol m⁻² and 1.28 μmol m⁻² for the adsorption of oxalate and citrate respectively are in good agreement with previously reported values for these species [31, 45]. Applying oxalic $\kappa'^{-1}$ values in Eq. 3a together with the $K_{fit}$ values allows the evaluation of $a_s$ values of 12.5, 12.6, 13.3, 19.6, 29.6 and 102.2 $m^2g^{-1}$ respectively for materials P1 to P6, while the citrate adsorption thus yields values of 3.9, 7.2, 10.6, 15.8, 44.2 and 100.1 $m^2g^{-1}$ for P1 to P6. It is important to note that these values are provisional as they are calculated on the basis of a linear fitting with relation to BET values. Extracting more definitive or meaningful numerical quantities would require the use of one or more reference samples of well characterised surface structure.

The computational and experimental methods followed here, in regards to the analysis of $\zeta$ variation as a function of mass-relative reagent addition with parameter fitting to the Grahame-Langmuir form to facilitate surface area assessment, can be applied with the use of a range of alternative complex cationic and anionic species potentially including, amongst others, the use of ammonium salts, phosphates/sulphates, xanthates or polysorbates. The appropriateness of these surface-interacting reagents is coupled to various substrate and system characteristics and necessitates adaptation on a case-by case basis. Importantly, electrokinetic parameters, determined here by PALS analysis of electrophoresis, can also be measured using alternative techniques including electro-osmosis, streaming potential or streaming current. Such methods may be more appropriate for conducting in-situ analyses of surface structure in more condensed particle-fluid systems such as slurries. This approach has relevance for the control and optimisation of industrial processes where the ability to assess surface structure of aqueous particles without lengthy ex-situ analysis is of value.

## 5. Conclusions

We have shown a numerical and experimental framework that can be applied to acquire surface structure information through means of electrokinetic characterisation. Specifically, the variation of zeta potential exhibited by adsorbent particles as a function of proportional adsorbate concentration follows the Grahame-Langmuir relationship for the monolayer-type adsorption of complex ionic species in aqueous media. The adsorption of carboxylate ions to alumina particles was used to demonstrate the merit of this behaviour for the quantitative and comparative characterisation of particle surfaces. A near linear relationship between the optimised value found from reverse-problem solving for the surface-area normalised adsorption coefficient $K$ and the conventionally determined surface area indicates that the analysis of zeta potential variation (or other electrokinetic properties) in conjunction with incremental addition of cationic or anionic adsorbates is applicable for the controllable scale specific evaluation of accessible surface area. By further employing ionic species of known and controlled adsorption cross section with respect to the substrate material in question, this approach can be used in the assessment of scale variance of fractal surface structures in aqueous particulate matter.

Using an automated dispension-measurement system, the experiments carried out here involved *in situ* electrokinetic analysis based surface characterisation. Thus, PALS based measurements and suspension modification were performed on a single recirculating aqueous system. This approach is advantageous relative to the analysis of gas accessible surface using $N_2$ at 77 K, as it allows application relevant room temperature analysis for aqueous particle-fluid systems in applications including water treatment, photocatalysis and industrial processing. Furthermore, such methods offer advantages in terms of rapidity and scalability relative to existing aqueous methods for surface area analysis which typically involve intermittent secondary analysis to assess levels of adsorption density or concentration of adsorbate in supernatant fluids [46].

The broad range of acceptable ionic adsorbate compounds that can be employed to facilitate the measurement of surface area by the assessment, *in situ* or otherwise, of indicative electrokinetic or electrochemical parameters is in contrast to existing methods for surface characterisation by adsorption isotherm interpretation that are inflexible with respect to surface-interacting compounds and thus do not readily facilitate application-specific analyses and the assessment of scale variance. Although precise quantitative evaluation of surface area requires calibration with reference to specimen of known surface area to facilitate back calculation, the use of in-situ electrokinetic analyses can readily facilitate the comparative analysis between suspended powders of varying surface structures, at tuneable scales.